\documentclass[twocolumn,showpacs,preprintnumbers,amsmath,amssymb,floatfix]{revtex4}
\usepackage{graphicx}
\usepackage{dcolumn}
\usepackage{bm}
\usepackage{color}
\definecolor{gold}{rgb}{0.85,.66,0}

\begin{document}

\title{Optical orientation of hole magnetic polarons in (Cd,Mn)Te/(Cd,Mn,Mg)Te quantum wells}

\author{E.~A.~Zhukov$^{1,2}$, Yu.~G.~Kusrayev$^2$, K.~V.~Kavokin$^{2,3}$, D.~R.~Yakovlev$^{1,2}$,
J.~Debus$^1$, A.~Schwan$^1$, I.~A.~Akimov$^{1,2}$, G.~Karczewski$^4$, T.~Wojtowicz$^4$,
J.~Kossut$^4$, and M.~Bayer$^{1,2}$}

\affiliation{$^1$ Experimentelle Physik 2, Technische Universit\"at Dortmund, 44221 Dortmund, Germany}
\affiliation{$^2$ Ioffe Institute, Russian Academy of Sciences, 194021 St.~Petersburg, Russia}
\affiliation{$^3$ Spin Optics Laboratory, St. Petersburg State University, 198504 St. Petersburg,
Russia}
\affiliation{$^4$ Institute of Physics, Polish Academy of Sciences, 02668 Warsaw, Poland}
\date{\today}

\begin{abstract}
The optically induced spin polarization in (Cd,Mn)Te/(Cd,Mn,Mg)Te
diluted-magnetic-semiconductor quantum wells is investigated by
means of picosecond pump-probe Kerr rotation. At 1.8~K temperature,
additionally to the oscillatory signals from photoexcited electrons
and Manganese spins precessing about an external magnetic field, a
surprisingly long-lived (up to 60~ns) nonoscillating spin
polarization is detected. This polarization is related to optical
orientation of equilibrium magnetic polarons involving resident
holes. The suggested mechanism for the optical orientation of the
equilibrium magnetic polarons indicates that the detected polaron
dynamics originates from unexcited magnetic polarons. The polaron
spin dynamics is controlled by the anisotropic spin structure of the
heavy-hole resulting in a freezing of the polaron magnetic moment in
one of the two stable states oriented along the structure growth
axis. Spin relaxation between these states is prohibited by a
potential barrier, which depends on temperature and magnetic field.
The magnetic polaron relaxation is accelerated with increasing
temperature and in magnetic field.
\end{abstract}

\pacs{78.55.Et, 73.21.Fg, 75.75.-c, 72.25.Fe }

\maketitle

\section{Introduction}
\label{sec:1}

Spin orientation and control in semiconductors and semiconductor
nanostructures are among the central tasks of spintronics. An
important related problem is the search for objects with long times
of spin coherence and relaxation and their investigation, which may
be also of interest for quantum information techniques
\cite{Awsch,Gaby_book}. All-optical experimental techniques are
often used for such investigations, in particular because they can
address the ultrafast spin dynamics of carriers, magnetic ions, and
magnetic excitations with picosecond and femtosecond time
resolution.

Optical orientation by circularly polarized light is a convenient
method for generating carrier spin polarization in semiconductors,
which can be transferred from the carriers to the localized spins of
nuclei and/or magnetic ions \cite{Opt-Orient}. For photoexcited
carriers, the spin lifetime depends on spin relaxation and
recombination. The latter provides a natural limit for the spin
lifetime of nonequilibrium carriers. To that end, the employment of
equilibrium carriers is preferable because their spin lifetime is
solely limited by spin relaxation and, therefore, can be long in case
of suppressed spin relaxation mechanisms.

Diluted magnetic semiconductors (DMS) based on II-VI materials, like
(Cd,Mn)Te and (Zn,Mn)Se, have often been used as model system to
study the spin dynamics of carriers and magnetic Mn$^{2+}$ ions
coupled by the strong \emph{s/p-d} exchange interaction. The interaction
results in fast spin relaxation of electrons and holes during a few
picoseconds \cite{Crooker96,Crooker97,Akimoto98}. Therefore, the
degree of spin polarization of nonequilibrium carriers is very low,
it does not exceed 1\%. However, the carriers can transfer their angular
momentum to the magnetic ions \cite{Krenn89,Crooker10,Akimov11}, whose
spin relaxation time may be considerably longer \cite{Yakovlev10b}.
Also the interaction with magnetic ions can stabilize the spin
orientation of localized carriers. This has been demonstrated by
optical orientation of exciton magnetic polarons (EMP)
\cite{Warnock1, Warnock2, Zakhar89, Merkulov95}, which are bound
complexes consisting of an exciton and about hundred Mn spins inside
the volume of the exciton localization.

Optical orientation of EMPs has been achieved only for selective
excitation in the spectral wing of the localized excitons. The induced
spin polarization is caused by thermodynamic fluctuations of the
magnetization inside the localization volume, providing a new
mechanism of optical orientation for DMS~\cite{Warnock1}.
The spin relaxation time of a magnetic polaron can be very long, but
the EMP recombines during about a nanosecond. Obviously, equilibrium
magnetic polarons involving resident carriers are much more
prospective candidates for achieving a long-lived spin polarization. This can
be realized for donor-bound or acceptor-bound magnetic polarons in
bulk DMS \cite{Wolff88}, or for magnetic polarons formed from
resident carriers localized in quantum wells or quantum dots,
while corresponding experimental observations have not been reported
so far. For investigating long-lived spin polarization, the
time-resolved pump-probe Faraday or Kerr rotation technique is more suitable
than photoluminescence spectroscopy, as the latter is naturally restricted
by the recombination time \cite{Crooker10}.

In this paper we report on optically induced polarization of hole
magnetic polarons (HMP) in (Cd,Mn)Te/(Cd,Mn,Mg)Te quantum wells,
examined by picosecond pump-probe Kerr rotation. In a transverse
magnetic field we find a nonoscillating component in the Kerr
rotation signal with a long decay time ($\leq$ 60~ns). We attribute
this signal to equilibrium magnetic polarons formed by resident
holes. A model for the optical orientation of HMPs is suggested and
details of its spin relaxation are analyzed theoretically and
studied experimentally.

The paper is organized as follows. In Section~\ref{sec:2} the
samples and experimental techniques are described. In
Section~\ref{sec:2b} the experimental data of the time-resolved
pump-probe Kerr rotation study on the spin dynamics in
(Cd,Mn)Te/(Cd,Mn,Mg)Te quantum wells are presented. The theoretical
model for the mechanism of optical orientation of the equilibrium
hole magnetic polarons is developed in Sec.~\ref{sec:3}. In
Section~\ref{sec:4} the spin relaxation of the
anisotropic hole magnetic polaron is described. The modeling of the
experimental data and the evaluation of the polaron parameters are
discussed in Sec.~\ref{sec:5}.

\section{Experimentals}
\label{sec:2}

Two Cd$_{1-x}$Mn$_x$Te/Cd$_{0.8-x}$Mn$_x$Mg$_{0.20}$Te  quantum well
(QW) structures were grown on $(100)$-oriented GaAs substrates by
molecular-beam epitaxy. Each structure consists of three
Cd$_{1-x}$Mn$_x$Te QWs with widths of $L=4$, 6, and 10~nm, separated
by 30-nm-thick Cd$_{0.8-x}$Mn$_x$Mg$_{0.20}$Te barriers. The Mn
concentration is $x=0.02$ (sample 041300A) and 0.04 (sample
041700В). The structures are nominally undoped, but the QWs contain
low concentrations of resident holes. This is confirmed
experimentally by the pump-probe measurements without and with
above-barrier illumination (see inset of Fig.~\ref{fig:B}), which
allows one to tune the concentration of resident carriers in the QWs
and even change the type of the majority background carriers~\cite{Zh09}.

\begin{figure}[t]
\includegraphics*[width=8 cm]{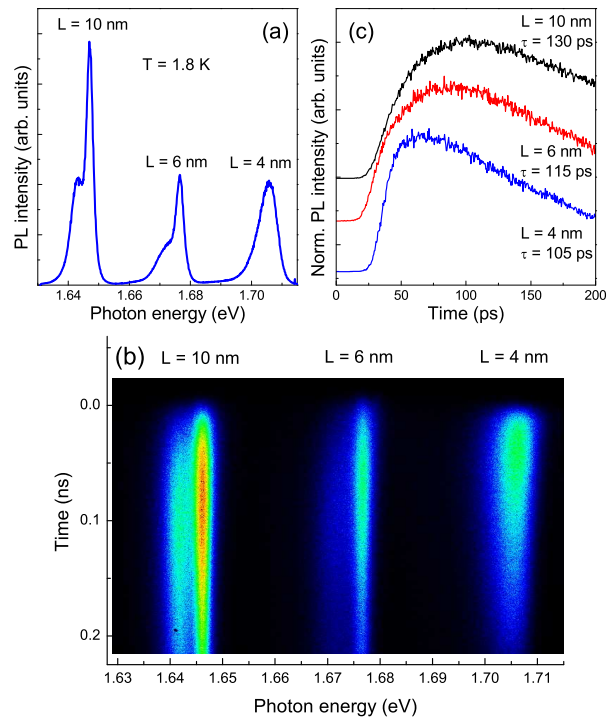}
\caption[] {(Color online) (a) Photoluminescence spectrum of the QWs
in the Cd$_{0.96}$Mn$_{0.04}$Te/Cd$_{0.76}$Mn$_{0.04}$Mg$_{0.20}$Te
heterostructure measured at $T=1.8$~K. (b) Streak-camera image of
the recombination dynamics excited at 1.718~eV, excitation density is 1~W/cm$^2$. (c) Recombination
dynamics of excitons and trions in the different QWs. }
\label{fig:A}
\end{figure}

The time-integrated photoluminescence (PL) spectrum of the sample with $x=0.04$
measured under pulsed excitation (same as streak-camera image described below) with photon energy of 1.718~eV at $T=1.8$~K is shown in  Fig.~\ref{fig:A}(a). The emission lines are composed of localized excitons and positively charged excitons (T$^+$ trions). They are well resolved in the 6- and 10-nm-thick QWs, where the trion line is a lower energy. In the 4-nm-thick inhomogeneous broadening due to well width fluctuations exceeds the trion binding energy and the exciton and trion lines are not resolved from each other.

The recombination dynamics of excitons and trions were measured by
time-resolved PL. The samples were excited by a pulsed Ti:Sapphire laser
(pulse duration 1.5~ps, repetition rate 76~MHz, photon energy
1.718~eV) and the emission was detected by a streak-camera attached
to a 0.5-m monochromator (time resolution of about 10~ps). The
recombination times $\tau$ of the QW exciton and trion in both
structures are in the range of 100~ps and do not exceed 130~ps at
$T=1.8$~K (see Figs.~\ref{fig:A}(b) and \ref{fig:A}(c)), which is typical for CdTe and (Cd,Mn)Te based QWs \cite{Yakovlev96,Yakovlev08}. Despite the presence of the magnetic
Mn ions, the dynamic spectral shift of the PL lines to low energies, which
would evidence an exciton magnetic polaron formation
\cite{Yakovlev10}, could not been detected. This is not surprising
because, for such low Mn concentrations, small EMP energies ($\le
1$~meV) are expected, so that the EMP formation time can exceed the
exciton recombination time. Both factors prevent EMP observation by
this experimental technique.

Time-resolved pump-probe Kerr rotation (KR) was used to study the
spin dynamics in these structures~\cite{Zh07,Crooker10}. Pulses with
a duration of 1.5~ps were taken from a Ti:Sapphire laser with a repetition
period $T_{\text{R}}= 13.2$~ns, which can be extended by a pulse-picker to
26.4, 52.9, 132, or 264~ns in order to analyze long-living
processes. The pump pulses were circularly polarized by an
elasto-optical modulator operated at 50~kHz frequency and their
energy was tuned either to the trion or exciton resonance. The
probe pulses were linearly polarized and their Kerr rotation was
measured by a balanced photoreceiver. For some measurements, a weak
above-barrier illumination with a photon energy of 2.34~eV was used to
tune the concentration and type of the resident carriers in the
QWs~\cite{Zh09,Deb13}. An external magnetic field $B$ of up to 2~T was
applied perpendicular to the QW growth $z$ axis (Voigt geometry).
The sample temperature was varied from $T=1.45$ to 10~K.

\section{Experimental results}
\label{sec:2b}

\begin{figure}[hbt]
\includegraphics*[width=8 cm]{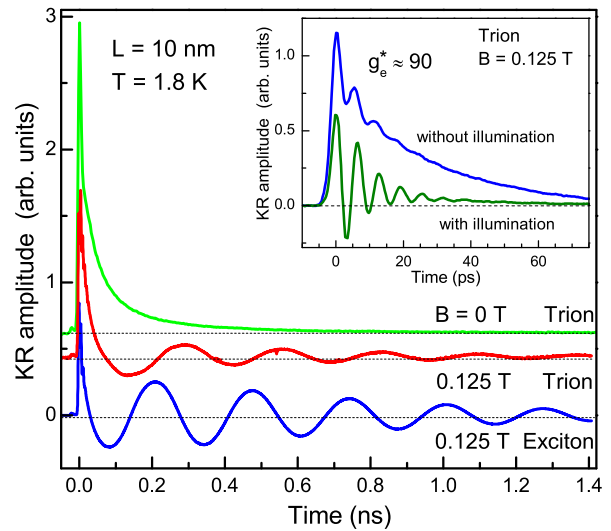}
\caption[] {(Color online) Kerr rotation signals of the 10-nm-thick
Cd$_{0.96}$Mn$_{0.04}$Te/Cd$_{0.76}$Mn$_{0.04}$Mg$_{0.20}$Te QW
measured on the trion resonance at $B=0$ and 0.125~T, and on the
exciton resonance at 0.125~T; $T_{\text{R}}=52.9$~ns. Inset:
electron spin beats at short time delays for $B=0.125$~T. Signals
without and with above-barrier illumination are shown. }
\label{fig:B}
\end{figure}

Pump-probe KR signals measured on the exciton and trion resonances
of the 10~nm QW with $x=0.04$ are shown in Fig.~\ref{fig:B}. It
shows many typical features of DMS QWs, compare, e.g.,
Refs.~\onlinecite{Crooker97} and \onlinecite{Crooker10}. At zero magnetic field the KR
amplitude has a fast and a slow decay component with times of 45~ps and
350~ps, respectively. At $B=0.125$~T the signal consists of three
components. Two of them are fast decaying, they are shown in the inset
of Fig.~\ref{fig:B}. The oscillating component arises from electron
spin beats with an effective $g$ factor $g^*\cong90$ and a dephasing
time of about 10~ps. This $g^*$ value corresponds to the slope of
the electron giant Zeeman splitting $d \Delta
E^{\mathrm{e}}/dB=5.4$~meV/T, which is in agreement with $x=0.04$
according to the growth parameters. As the hole exchange constant in
(Cd,Mn)Te is four times larger than that of the electron, one can evaluate
the magnetic field slope of the heavy-hole giant Zeeman splitting as
$d \Delta E^{\mathrm{hh}}_z/dB=21.6$~meV/T. The nonoscillating
component in the inset is attributed to resident heavy holes with
strongly anisotropic $g$ factor with an in-plane component close to
zero. The relative amplitudes of the electron and hole contributions
are changed when the sample is exposed to additional above barrier
illumination with a density of 0.3~W/cm$^2$ that enhances the
electron contribution. It was shown for CdTe/(Cd,Mg)Te QWs that at
low temperatures the electrons photoexcited by above-barrier
excitation are predominantly captured by the QWs, while the holes
are partly localized in the barriers by alloy
fluctuations~\cite{Zh09}. In the studied structure this decreases
the concentration of resident holes. The experimental results in the
inset of Fig.~\ref{fig:B} let us conclude that the studied samples
contain resident holes in the QWs.

\begin{figure}[hbt]
\includegraphics*[width=8 cm]{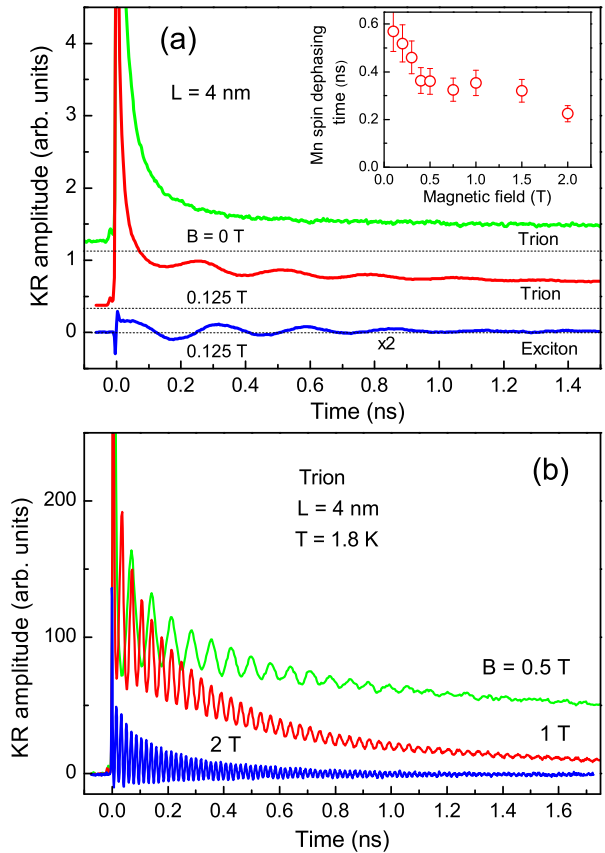}
\caption[] {(Color online) Kerr rotation signals of the 4-nm-thick
Cd$_{0.96}$Mn$_{0.04}$Te/Cd$_{0.76}$Mn$_{0.04}$Mg$_{0.20}$Te QW;
$T_{\text{R}}=52.9$~ns. (a) Curves measured on the trion at $B=0$
and on the trion as well as exciton at $B=0.125$~T are shifted
vertically for clarity; their zero levels are shown by dashed lines. Inset: Spin dephasing time of the Mn
spins as function of magnetic field. (b) KR signals measured on the trion at different magnetic fields.
Note that all spectra have the same zero level. The nonoscillating
component is very pronounced here. }
\label{fig:1}
\end{figure}

The slow precession for delay times of up to 1.4~ns in Fig.~\ref{fig:B}
is attributed to the Mn$^{2+}$ spins with $g$ factor
$g_{\mathrm{Mn}}=2.0$ and dephasing time of 530~ps. The dephasing
time decreases to 350~ps at $B=0.5$~T and then stays about constant
up to 2~T, see inset of Fig.~\ref{fig:1}(a). It is worthwhile mentioning, that the signals measured
on the exciton and trion resonances are quite similar to each other.
Similar KR signals are also obtained for the 6-nm-thick QW and the 6-
and 10-nm-thick QWs with $x=0.02$.

The KR signal measured on the trion in the 4-nm-wide QW with $x=0.04$
shows a nonoscillatory component with large amplitude and long decay
time, see Fig.~\ref{fig:1}(a). The decay at $B=0$~T exceeds the
laser pulse separation $T_{\text{R}}=52.9$~ns, as we detect a finite KR
amplitude at negative time delays. This component is absent in the
signal detected at the exciton resonance. Signals for several
magnetic fields are presented in Fig.~\ref{fig:1}(b). Here, the
decay time $\tau_{\mathrm{MP}}$ of the nonoscillating component
measured on the trion is $\simeq 3$~ns at $B=0.5$~T. The decay time
and amplitude of this component strongly decrease with further
increasing magnetic field up to 2~T.

\begin{figure}[hbt]
\includegraphics*[width=8 cm]{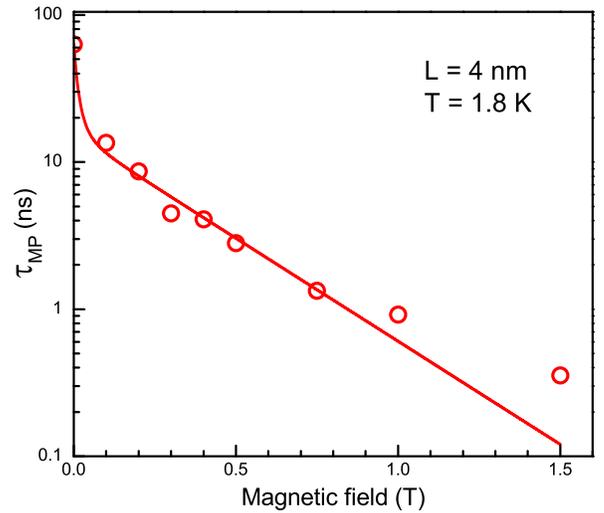}
\caption[] {(Color online) Magnetic field dependence of the spin
relaxation time $\tau_{\mathrm{MP}}$ in the
Cd$_{0.96}$Mn$_{0.04}$Te/Cd$_{0.76}$Mn$_{0.04}$Mg$_{0.20}$Te QW with
$L=4$~nm; $T_{\text{R}}=52.9$~ns, $T=1.8$~K. Experimental data are shown
by symbols. The line is a fit to the data using Eqs.~\eqref{t12} and
\eqref{t10} with the fitting parameters: $B_{\mathrm{ex}}=0.05$~T
and $g_{\bot}/g_{zz}=0.04$. } \label{fig:4}
\end{figure}

The magnetic field dependence of the decay time of the
nonoscillating component is shown in Fig.~\ref{fig:4}. We denote
this time by $\tau_{\mathrm{MP}}$, as it is attributed to the directional relaxation of the magnetic moment of the magnetic polaron,
i.e., to the magnetic-polaron spin relaxation; it will be shown in the following. One can see that $\tau_{\mathrm{MP}}$
decreases from 60~ns at zero field down to 0.35~ns at $B=1.5$~T. At
zero field $\tau_{\mathrm{MP}}$ was evaluated from the dependence of
the KR amplitude at a small negative time delay of 60~ps prior to
the pump pulse on the temporal pump-pulse separation $T_{\text{R}}$, which was varied
from 13.2 to 264~ns. The dependence in Fig.~\ref{fig:4} can be described by two
different ranges: one at low magnetic fields up to 0.1~T, where a
fast decrease of $\tau_{\mathrm{MP}}$ occurs, and a second one at
higher fields, where the decay becomes much weaker. This evidences
that there are two different regimes for the HMP dynamics at low and
high magnetic fields, whose features will be discussed in
Sec.~\ref{sec:4}.

\begin{figure}[hbt]
\includegraphics*[width=8 cm]{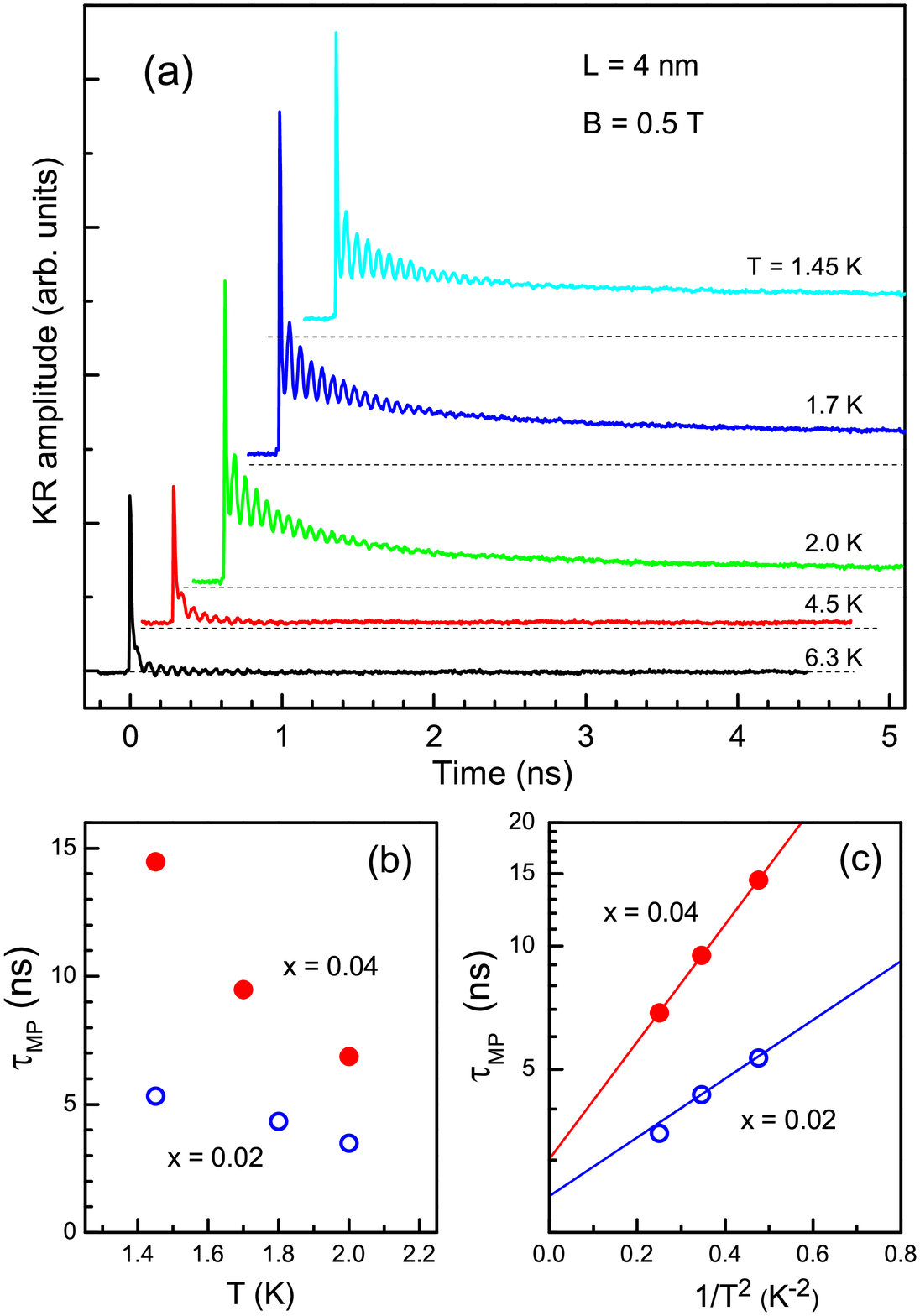}
\caption[] {(Color online)  (a) Temperature dependence of the Kerr rotation signal recorded on
the trion resonance of the 4-nm-thick Cd$_{0.96}$Mn$_{0.04}$Te/Cd$_{0.76}$Mn$_{0.04}$Mg$_{0.20}$Te QW;
$T_{\text{R}}=13.2$~ns and pump power is 1~W/cm$^2$. Zero levels are shown by dashed lines. The signals are
offset in time to avoid their overlap. (b) Temperature dependence of
the spin relaxation time of the nonoscillating component attributed
to the HMP spin relaxation time $\tau_{\mathrm{MP}}$. (c) Logarithmic plot of
the spin relaxation rate of the nonoscillating component as function of $1/T^2$. } \label{fig:2}
\end{figure}

Figure~\ref{fig:2}(a) presents the KR signals of the 4-nm-thick QW
measured at different temperatures.  With increasing temperature
from 1.45 to 6.3~K the decay time of the nonoscillating component
decreases drastically to a value $\tau_{\mathrm{MP}}$ below 100~ps.
The temperature dependence of $\tau_{\mathrm{MP}}$ is shown in
Fig.~\ref{fig:2}(b). The time shortening is accompanied with a
pronounced amplitude decrease.

\begin{figure}[hbt]
\includegraphics*[width=8 cm]{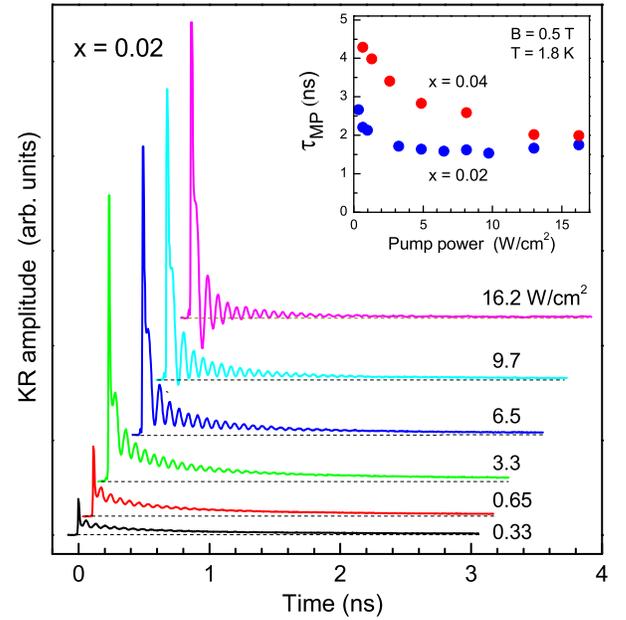}
\caption[] {(Color online) (a) Pump power dependence of the Kerr rotation signal recorded on the
trion resonance of the 4-nm-thick Cd$_{0.98}$Mn$_{0.02}$Te/Cd$_{0.78}$Mn$_{0.02}$Mg$_{0.20}$Te QW; $T_{\text{R}}=13.2$~ns, $T=1.8$~K. Zero levels are
shown by dashed lines. The signals are offset in time to avoid their
overlap. Inset: pump power dependences of the decay time
of the nonoscillating component for the 4-nm-thick QWs with $x=0.02$ and
0.04. } \label{fig:C}
\end{figure}

The sample with $x=0.02$ shows  a behavior qualitatively similar to
the results of the sample with $x=0.04$. In Fig.~\ref{fig:C} we show the impact of changes in the pump power on the KR signals from the 4-nm-thick QW with $x=0.02$. With increasing power the nonoscillating component disappears, its amplitude decreases and the decay time becomes shorter, see also inset of Fig.~\ref{fig:C}. In this respect the behavior is similar to that induced by a temperature increase, compare with Fig.~\ref{fig:2}(a).

The long-lived nonoscillating signal observed in the Voigt geometry
for the applied external magnetic fields deserves special attention,
as it has not been reported earlier for DMS QWs. It cannot originate from optically oriented electrons or holes, as the carrier
spin relaxation times in DMS structures typically do not exceed tens
of picoseconds \cite{Crooker97,Akimoto98,Camilleri01,Freeman90}. It
also unlikely that it is attributed to the dynamics of Mn$^{2+}$ spins, whose
precession is clearly observed with an amplitude that weakly depends
on temperature and magnetic field. The unusual properties can be
related only to a hole magnetic polaron, for which a strong spin
anisotropy, namely a small in-plane $g$ factor, is provided by the
strongly confined heavy hole, whose spin relaxation becomes locked
through the interaction with Mn spins. In the following sections we will
analyze the mechanism of the optical orientation of the HMP and the
dependence of its spin relaxation time on the temperature and magnetic
field strength.

\section{Model of optical orientation of hole magnetic polaron}
\label{sec:3}

The observed experimental data can be explained by the optical
orientation of equilibrium magnetic polarons involving resident
holes. In QWs, the heavy-hole and light-hole states with angular
momentum projections $J_z=\pm 3/2$ and $\pm 1/2$, respectively, onto
the structure growth axis ($z$ axis) are split by lattice strain and
quantum confinement. The lowest heavy-hole state is described as a
quasi-particle with strongly anisotropic $g$-factor tensor: $g_{xx},
g_{yy} \ll g_{zz}$ \cite{Merk95}. A magnetic polaron, which is formed by a
heavy hole interacting with Mn spins, therefore also exhibits a
strong anisotropy. At zero magnetic field, the polaron formation leads to
maximum polaron binding energy (or magnetic polaron energy)
$E_{\mathrm{MP}} = M_{\mathrm{MP}}B_{\mathrm{ex}}$, when the magnetic
moment of the polaron $\mathbf{M}_{\mathrm{MP}}$ (the total magnetic
moment of the Mn ions inside the hole localization volume) is
directed along the $z$ axis. Here, $B_{\mathrm{ex}}$ is the mean
exchange field of the localized hole acting on the Mn spins. For
spin relaxation, the HMP should flip its magnetic moment and
overcome the energy barrier equal to $E_{\mathrm{MP}}$ between two
stable states along the $z$ axis.

Let us discuss the mechanism of optical generation of spin
polarization for an ensemble of equilibrium HMPs localized in a QW.
Several aspects should be taken into account: (i) The HMPs involving
equilibrium resident holes are in equilibrium, i.e., they have
infinite lifetime and their binding energy reaches the value
$E_{\mathrm{MP}}$. (ii) The HMPs are formed from the anisotropic
heavy-holes with $J_z = \pm 3/2$ \cite{Comm1}. (iii) The circular-polarized
photoexcitation is resonant with the lowest possible state that, for
the HMP, is the positively charged exciton $T^+$ consisting of two
holes in the spin singlet state and an electron \cite{Bartsch11}. In
this respect the suggested model has some analogy with the model
describing the generation of spin coherence in low-density electron
or hole gases via resonant excitation of trion states in QWs
\cite{Zh07,Yakovlev08}.

\begin{figure}[hbt]
\includegraphics*[width=8 cm]{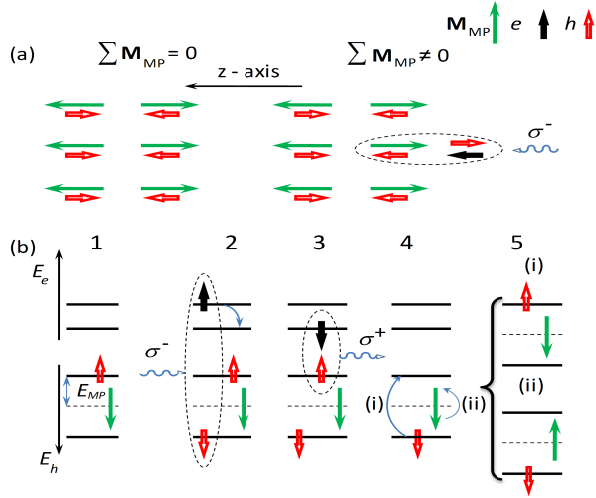}
\caption[] {(Color online) Mechanism of optical orientation of an
ensemble of HMPs. (a) Trion photoexcitation with a $\sigma^-$
polarized photon depolarizes the HMP with specific orientation. (b)
Details of HMP depolarization (see text). Red and black arrows
represent the spins of heavy-hole and electron, respectively. Green
arrow does the same for $\mathbf{M}_{\mathrm{MP}}$; $B=0$~T. }
\label{fig:5}
\end{figure}

At zero magnetic field the magnetic moments of the HMPs,
$\mathbf{M}_{\mathrm{MP}}$, are equally distributed between the
orientations parallel and antiparallel to the $z$ axis so that the
total polarization of the HMP ensemble is zero, see
Fig.~\ref{fig:5}(a). The resonant excitation of the trion states by,
e.g., $\sigma^-$ circularly polarized light photocreates T$^+$ trions from
the resident holes, whose spin is oriented parallel to the $z$ axis. As a result,
the HMPs formed by these holes become heated (excited) and fully or partly
depolarized. This induces a net spin orientation
in the ensemble of HMPs. For pump-probe KR detection the signal is
contributed by the unperturbed HMPs with $\mathbf{M}_{\mathrm{MP}}$
parallel to the $z$ axis. It is interesting that in this case we are
thus able to study the properties of equilibrium HMPs, which are not
affected by photoexcitation. This is different from common optical
orientation experiments, where the polarized photoluminescence of
photoexcited spin-oriented carriers is analyzed.

Let us discuss in detail the depolarization mechanism of
photoexcited HMPs. The respective schematic diagrams are shown in
Fig.~\ref{fig:5}(b). Stage 1 corresponds to the equilibrium HMP,
here the hole spin is antiparallel to $\mathbf{M}_{\mathrm{MP}}$. It
is important to note, that the spin splitting of the hole and
electron states is not induced by an external magnetic field, which
is zero, but is due to carrier exchange interaction with the polaron
magnetic moment. The photoexcitation of the trion ground state from this
resident hole with $\sigma^-$ circularly polarized light is shown as
stage 2. Here, a hole with spin-down is generated (spin-up state is
already occupied by the resident hole) and an electron with spin-up.
For the electron, this is an energetically unfavorable orientation
and it relaxes into the spin-down state. Then, the spin-down
electron recombines with the spin-up hole resulting in emission of a
$\sigma^+$ polarized photon (stage 3) and leaving the resident
spin-down hole in an orientation unfavorable for
$\mathbf{M}_{\mathrm{MP}}$ (stage 4). There are two ways to obtain
again the favorable orientation: (i) to flip the hole spin and end up
with an orientation identical to the initial one (stage 1), or (ii)
to flip $\mathbf{M}_{\mathrm{MP}}$, which will change the order of
the spin levels for holes and provide a HMP with an orientation
opposite to the initial one. As a result, the subensemble of spin-up
oriented HMPs becomes partly depolarized, which in turn leads to optical
orientation of the whole ensemble of HMPs. Note, that both electron
and hole spin relaxation in presence of $\mathbf{M}_{\mathrm{MP}}$
requires a dissipation of energy, which is typically transferred to
phonons or the Mn spin reservoir and, therefore, heats the Mn spin
system inside the HMP. This heating may induce a reorientation of
$\mathbf{M}_{\mathrm{MP}}$.

\section{Model of spin relaxation of anisotropic hole magnetic polaron}
\label{sec:4}

\begin{figure}[hbt]
\includegraphics*[width=8 cm]{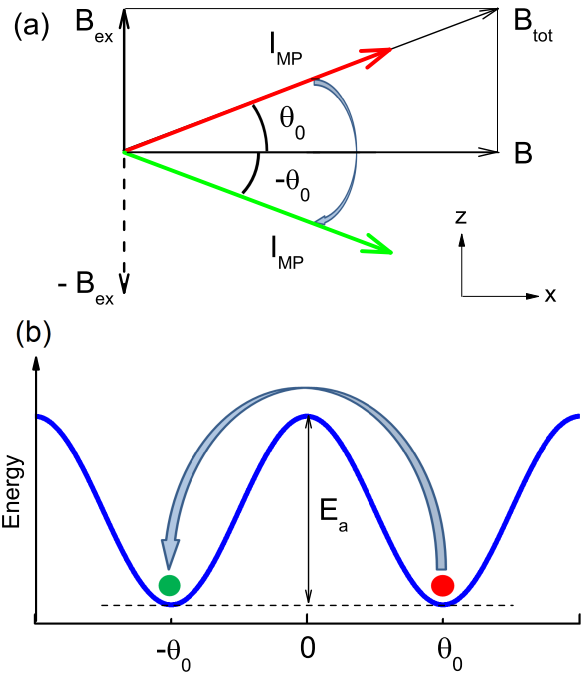}
\caption[] {(Color online) (a) Scheme of spin structure of the HMP in an
external magnetic field applied in Voigt geometry. (b) Schematic
presentation of the HMP potential. The spin relaxation of the HMP corresponds to a transition between
two states with minimal energy, separated by the potential barrier
$E_a$. The figure is valid for a strongly anisotropic heavy-hole spin struture in a
DMS QW.}
\label{fig:7}
\end{figure}

The excited optical orientation of the HMPs relaxes with the polaron
spin relaxation time $\tau_{\mathrm{MP}}$. It is important that due
to their anisotropic spin structure the HMPs do not participate in
coherent spin precession about an external magnetic field, but
rather show up as decaying contribution to the pump-probe Kerr
rotation signal. To flip $\mathbf{M}_{\mathrm{MP}}$, i.e., to provide
polaron spin relaxation, it is necessary to overcome a barrier with an
activation energy $E_{\mathrm{a}}$, as shown schematically in
Fig.~\ref{fig:7}(b). At zero external magnetic field
$E_{\mathrm{a}}$ is equal to $E_{\mathrm{MP}}$, but the barrier decreases at a
finite magnetic field strength. At low temperatures ($E_{\mathrm{a}} \gg
k_\mathrm{B}T$) the HMP spin relaxation can last very long, in the
studied structures up to 60~ns. With growing temperature the spin-flip
probability of the polaron magnetic moment increases. The strong
temperature dependence of the HMP spin-relaxation time observed in
the experiment suggests that it occurs due to an activation process,
which can be described by
\begin{equation}
\tau_{\mathrm{MP}}(B,T) \approx \tau_0 \exp \left( \frac{E_{\mathrm{a}}(B,T)}{k_\mathrm{B}T} \right).
\label{t12}
\end{equation}
Here, $\tau_0$ is a pre-factor of the order of the hole spin-flip
time. In this section we will model the magnetic field and
temperature dependences of $E_{\mathrm{a}}(B,T)$ to describe the
experimental observations and evaluate the HMP parameters.

To calculate the dependence of the activation energy on the external
magnetic field, we make use of the theory of an anisotropic magnetic
polaron formed by holes in quantum wells, developed by Merkulov and
Kavokin~\cite{Merk95}. The magnetic polaron (MP) possesses a huge
angular momentum mainly due to the total spin of many Mn ions within
the localization volume of the hole. The spin Hamiltonian of the
exchange-coupled Mn and hole-spins in a magnetic field $\mathbf{B}$,
\begin{eqnarray}
\hat{H}_S = -\frac{1}{3}\beta \mathbf{J}\cdot \sum_{i} \mathbf{S}_i  |\Psi(\mathbf{r}_i)|^{2} +
g_{\mathrm{Mn}}\mu_\mathrm{B}\mathbf{B}\cdot\sum_{i} \mathbf{S}_i \nonumber\\ +  \hat{H}_{Zh} +\hat{H}_{SO}     \label{Ham_A}
\end{eqnarray}
comprises four parts: the exchange interaction, the Zeeman
interaction of the Mn ions, the Zeeman interaction of the hole
$\hat{H}_{Zh}$ (the latter is usually neglected because it is much
smaller than the exchange interaction with the Mn ions), and the
spin-orbit interaction $\hat{H}_{SO}$ responsible for splitting the
heavy- and light-hole subbands in QWs. Here, $\beta$ is the
\emph{p-d} exchange constant, $\mathbf{J}$ is the hole spin,
$\mathbf{S}_i$ is the spin of the \emph{i}-th Mn ion located at the
the position vector $\mathbf{r}_i$, $\Psi(\mathbf{r})$ is the hole
envelope function, $g_{\mathrm{Mn}}=2.01$ is the $g$ factor of the
Mn$^{2+}$ ion, and $\mu_\mathrm{B}$ is the Bohr magneton. Numerous
previous studies have shown that the experimental manifestations of
the MP are well described by the "box" model, replacing the operator
$\sum_{i} \mathbf{S}_i  |\Psi(\mathbf{r}_i)|^{2}$ with the total
spin operator $\hat{\mathbf{I}}$ of the Mn ions within the magnetic
polaron volume that is defined by
\begin{eqnarray}
V_{\mathrm{MP}} = \left(\int |\Psi(\mathbf{r})|^{4} d^3r \right)^{-1}.
\label{MPvolume}
\end{eqnarray}
Within the "box" approximation, the Hamiltonian \eqref{Ham_A} takes a simplified form
\begin{eqnarray}
\hat{H}_S = -\frac{1}{3V_{\mathrm{MP}}}\beta \mathbf{J}\cdot \hat{\mathbf{I}} +
g_{\mathrm{Mn}}\mu_\mathrm{B}\mathbf{B}\cdot \hat{\mathbf{I}} +  \hat{H}_{SO}  .
\label{Ham_B}
\end{eqnarray}
If the energy splitting between the heavy holes and light holes is much larger than
the exchange energy of the hole (which is true in narrow quantum
wells), one can use a truncated Hamiltonian restricted to the
ground-state heavy-hole spin doublet. Merkulov and
Kavokin~\cite{Merk95} suggested writing it in the form
\begin{eqnarray}
\hat{H}_S = -\frac{1}{3V_{\mathrm{MP}}}\beta \sum_{k,l} \hat{j}_k g_{kl} \hat{I}_l +
g_{\mathrm{Mn}}\mu_\mathrm{B}\mathbf{B}\cdot \hat{\mathbf{I}}.
\label{Ham_C}
\end{eqnarray}
Here, $\hat{\mathbf{j}}$ is the pseudospin $1/2$ operator of the hole
with the components given by the Pauli matrices defined in the
heavy-hole $J_z=\pm3/2$ subspace (note that the exchange constant
$\alpha$ in Ref.~\onlinecite{Merk95} equals to $\beta /
3V_{\mathrm{MP}}$),  $g_{kl}$ is the tensor of the effective $g$
factor of the heavy hole (usually
$\left|g_{xx}\right|=\left|g_{yy}\right|=g_{\perp}\ll g_{zz}=3$),
and $k,l \in \{x,y,z\}$.

While at the initial stages of MP formation its energy may be
determined by fluctuations of the Mn total spin $\hat{\mathbf{I}}$,
once the MP is formed, the main contribution to the energy comes from
the interaction of the hole spin with the mean $\mathbf{I}$ vector
(collective regime). In this regime, the hole occupies the lowest
spin level in the effective field generated by the polarized Mn
spins, and the polaron energy can be written as
\begin{eqnarray}
E = -\frac{1}{3V_{\mathrm{MP}}} \beta \sum_{k,l}j_{k}g_{kl}I_{l} +g_{\mathrm{Mn}}\mu_\mathrm{B}\mathbf{B}\cdot \mathbf{I}.
\label{t1}
\end{eqnarray}
Here, the $I_{l}$ are the components of the mean total spin of the Mn
ions inside the polaron volume, and the $j_{k}$ are the components
of the mean hole pseudospin, which is a Bloch vector defining a
linear combination of states within the heavy-hole doublet. We
choose the \emph{z} axis to be normal to the QW plane,
i.e., parallel to the structure growth axis. We would like to stress
here that Eq.~\eqref{t1} describes not only the ground state of the
polaron, but also states with arbitrary direction of $I$. The
activation energy is equal to the energy difference between the
polaron ground state and the highest point of the transition path (see Fig.~\ref{fig:7}(b)).
Let us first calculate the energy for the polaron ground state as a
function of the applied magnetic field.

We consider the Voigt geometry, when the magnetic field is in the QW
plane and has only a component along the $x$ axis:
$\mathbf{B}=(B,0,0)$. Since for II-VI diluted magnetic
semiconductors the \emph{p-d} exchange constant is negative
($\beta<0$), in the polaron ground state the hole pseudospin is
anti-parallel to $\mathbf{I}$ and $E$ should be minimal. This
consideration leads to the following expressions for the components
of $\mathbf{j}$:
\begin{eqnarray}
j_{x}&=&-\frac{1}{2}\frac{I_{x}g_{\perp}}{\sqrt{I_{z}^{2}g_{zz}^{2}+I_{x}^{2}g_{\perp}^{2}}},\nonumber\\
j_{y}&=&0,  \\
j_{z}&=&-\frac{1}{2} \frac{I_{z}g_{zz}}{\sqrt{I_{z}^{2}g_{zz}^{2}+I_{x}^{2}g_{\perp}^{2}}}\nonumber.
\label{t2}
\end{eqnarray}
As a function of the angle $\theta$ between $\mathbf{I}$ and the
in-plane magnetic field $\mathbf{B}$, $j_{z}$ remains close to
$-1/2$, while $\pi/2\geq|\theta|>g_{\perp}/g_{zz}$. In this range of
angles the polaron energy is given by a simple expression:
\begin{equation}
E\left(\theta\right)=-\frac{|\beta|}{6V_{\mathrm{MP}}}g_{zz}I\mid\sin\theta\mid - g_{\mathrm{Mn}}\mu_{B}BI\cos\theta.
\label{t3}
\end{equation}
It reaches the minimum value at
$\theta_{0}=\arctan\left(\frac{g_{zz}\beta}{6g_{\mathrm{Mn}}\mu_\mathrm{B}B}\right)$,
which corresponds to the ground state of the polaron. The energy of
this state is
\begin{eqnarray}
E(\theta_{0})&=&-I\sqrt{\beta(g_{zz}/6V_{\mathrm{MP}})^{2}+(g_{\mathrm{Mn}}\mu_\mathrm{B}B)^{2}}\nonumber\\
&=&-g_{\mathrm{Mn}}\mu_\mathrm{B}I\sqrt{B_{\mathrm{ex}}^{2}+B^{2}},
\label{t4}
\end{eqnarray}
where
\begin{eqnarray}
B_{\mathrm{ex}}=\frac{|\beta| g_{zz}}{6g_{\mathrm{Mn}}\mu_\mathrm{B}V_{\mathrm{MP}}}
\label{t20}
\end{eqnarray}
is the exchange field generated by the hole (see Eq.~(7.12) of
Ref.~\onlinecite{Yakovlev10}, note that $g_{zz}=3$).
For simplicity, we will continue the considerations by suggesting a
linear dependence of the Mn polarization on $B$. This approach is
valid for the conditions of our experiment, where $B<2$~T and the
expected $B_{\mathrm{ex}}<2$~T. The magnitude
of $I$ in the polaron ground state, $I_{\mathrm{MP}}$, is determined
by the total field that is applied to the Mn ions within the
polaron, which is the vector sum of the external field and the
exchange field:
\begin{equation}
I_{\mathrm{MP}}=\frac{V_{\mathrm{MP}}\chi}{g_{\mathrm{Mn}}\mu_\mathrm{B}}\sqrt{B_{\mathrm{ex}}^{2}+B^{2}},
\label{t5}
\end{equation}
where $\chi$ is the magnetic susceptibility of the Mn ions. Note
that the magnitude of the HMP magnetic moment is
$M_{\mathrm{MP}}=g_{\mathrm{Mn}}\mu_\mathrm{B}I_{\mathrm{MP}}$.

The two ground states of the HMP in an external magnetic field are
shown schematically in Fig.~\ref{fig:7}(a). The blue arrow marks the
transition between these states, which provides the spin relaxation
of the HMP with the characteristic time $\tau_{\mathrm{MP}}$.

Combining Eqs.~\eqref{t4} and \eqref{t5}, we obtain
\begin{eqnarray}
E(\theta_{0})&=&-V_{\mathrm{MP}}\chi\left(B_{\mathrm{ex}}^{2}+B^{2}\right)\nonumber\\
&=& E\left(B=0\right)\left(1+\frac{B^{2}}{B_{\mathrm{ex}}^{2}}\right).
\label{t6}
\end{eqnarray}

Relaxation of the HMP magnetic moment is provided by changing the
orientation of $\mathbf{I}_{\mathrm{MP}}$ between its stable states
at $\pm \theta_0$, see Fig.~\ref{fig:7}(b). The highest point of the
most probable activation path of $\mathbf{I}_{\mathrm{MP}}$ is at
$\theta=0$ (where the polaron energy has a saddle point as function
of the direction of $\mathbf{I}$). At this point,

\begin{eqnarray}
E\left(\theta=0\right)&=&-g_{\mathrm{Mn}}\mu_\mathrm{B}BI_{\mathrm{MP}}-\frac{|\beta|}{6V_{\mathrm{MP}}}g_{\perp}I_{\mathrm{MP}}\nonumber\\
&=&-g_{\mathrm{Mn}}\mu_\mathrm{B}I_{\mathrm{MP}}\left(B+\frac{g_{\perp}}{g_{zz}}B_{\mathrm{ex}}\right)\nonumber\\
&=&-V_{\mathrm{MP}}\chi\sqrt{B_{\mathrm{ex}}^{2}+B^{2}}\left(B+\frac{g_{\perp}}{g_{zz}}B_{\mathrm{ex}}\right)\nonumber\\
&=&E\left(B=0\right)\left(\frac{B}{B_{\mathrm{ex}}}+\frac{g_{\perp}}{g_{zz}}\right)\sqrt{1+\frac{B^{2}}{B_{\mathrm{ex}}^{2}}}.
\label{t7}
\end{eqnarray}
The activation energy is given by
\begin{eqnarray}
E_{\mathrm{a}}=E\left(\theta=0\right)-E\left(\theta_{0}\right)\nonumber\\
=E\left(B=0\right) \left[\left(\frac{B}{B_{\mathrm{ex}}}+
\frac{g_{\perp}}{g_{zz}}\right)\sqrt{1+\frac{B^{2}}{B_{\mathrm{ex}}^{2}}}-\left(1+\frac{B^{2}}{B_{\mathrm{ex}}^{2}}\right)\right].
\label{t8}
\end{eqnarray}

Note that at zero external magnetic field the HMP is aligned along
the \emph{z} axis ($\theta_{0}=\pm\pi/2$), $I_{\mathrm{MP}}(B=0)=
V_{\mathrm{MP}} \chi B_{\mathrm{ex}} / (g_{\mathrm{Mn}}
\mu_\mathrm{B})$ and the polaron energy according to Eq.~\eqref{t4}
is given by
\begin{eqnarray}
E(B=0)=-V_{\mathrm{MP}} \chi B_{\mathrm{ex}}^{2} =-E_{\mathrm{MP}}(B=0).
\label{EMP2}
\end{eqnarray}
Here, $E_{\mathrm{MP}}(B=0)>0$ is the magnetic polaron binding energy, which has by definition a positive
value~\cite{Kavokin99b,Yakovlev10}.
In the linear approximation the expression for the HMP energy can be obtained using the definition of the exchange field (Eq.~\eqref{t20}):
\begin{equation}
E\left(B=0\right)=-\frac{1}{2}\frac{d \Delta E^{\mathrm{hh}}_{z}}{dB} B_{\mathrm{ex}}   ,
\label{t9}
\end{equation}
where $\Delta E^{\mathrm{hh}}_{z}=|\beta)|g_{zz}/[3I_z(B_z)]$ is the hole Zeeman splitting in Faraday geometry.

In this case we obtain
\begin{eqnarray}
E_{\mathrm{a}}= \frac{1}{2}\frac{d \Delta E^{\mathrm{hh}}_{z}}{dB}
B_{\mathrm{ex}}\sqrt{1+\frac{B^{2}}{B_{\mathrm{ex}}^{2}}}
\left(\sqrt{1+\frac{B^{2}}{B_{\mathrm{ex}}^{2}}}-
\frac{B}{B_{\mathrm{ex}}}-\frac{g_{\perp}}{g_{zz}}\right).
\label{t10}
\end{eqnarray}

\begin{figure}[hbt]
\includegraphics*[width=8 cm]{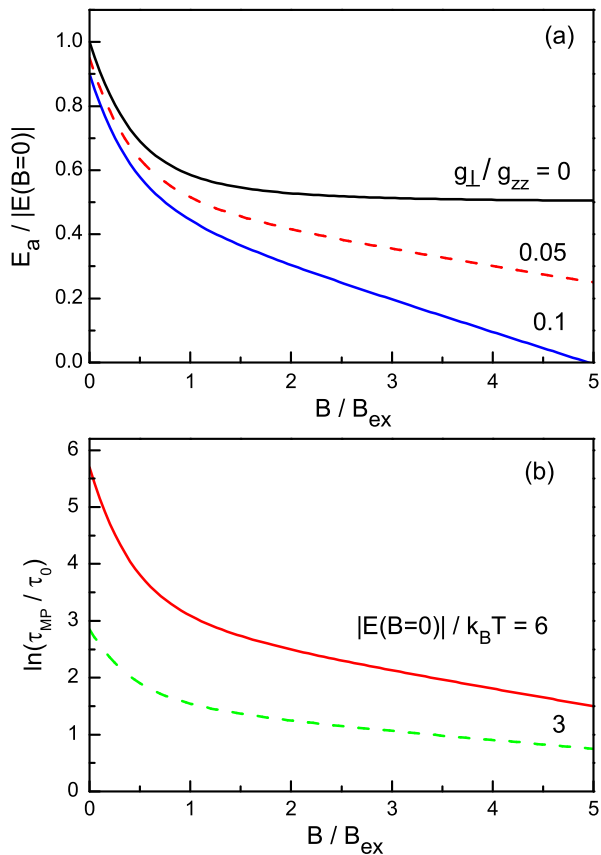}
\caption[] {(Color online) Calcualtion of the magnetic field
dependences of (a) the activation energy normalized on $E(B=0)$
using Eq.~\eqref{t10}, and (b) the HMP spin relaxation time
normalized on $\tau_0$ according to Eq.~\eqref{t12}. The magnetic
fields are give in units of $B_{\mathrm{ex}}$. } \label{fig:6}
\end{figure}

The magnetic field dependence of the activation energy, given by
Eq.~\eqref{t10}, can be split in two ranges. In weak external
fields, the ground-state polaron energy does not change
considerably, because the external field is applied perpendicular to
the hole exchange field. Therefore, the lowering of the activation
energy is determined by the change of the barrier height, equal to
the Zeeman shift of the lowest hole spin level:
\begin{equation}
E_{\mathrm{a}}\left(B\rightarrow0\right)\approx\frac{1}{2}\frac{d \Delta E^{\mathrm{hh}}_{z}}{dB}\left(B_{\mathrm{ex}}-B\right).
\label{t11}
\end{equation}
When the external field becomes comparable in strength with the hole
exchange field, the total energy of the polaron in its ground state
starts to decrease with increasing external field, thus compensating
the decrease of the barrier height. As a result, the lowering of the
activation energy is considerably weakened:
\begin{equation}
E_{\mathrm{a}}\left(B\rightarrow\infty\right)\approx\frac{1}{2}\frac{d \Delta E^{\mathrm{hh}}_{z}}{dB}\left(\frac{1}{2}B_{\mathrm{ex}}-\frac{g_{\bot}}{g_{zz}}B\right).
\label{t11a}
\end{equation}
The crossover between the two regimes occurs at
$B\approx\frac{1}{2}B_{\mathrm{ex}}$. As one can see from
Eq.~\eqref{t11a}, in the case of $g_{\bot}=0$ the field dependence
of the activation energy saturates at $E_{\mathrm{a}}(B=0)/2$. Its
slope for $B>\frac{1}{2}B_{\mathrm{ex}}$ gives the ratio of the
principal values of the hole $g$-factor tensor $g_{\bot}/g_{zz}$. These two
regimes are clearly seen in Fig.~\ref{fig:6}(a), where the
activation energy $E_{\mathrm{a}}$ normalized to the magnetic
polaron energy at zero magnetic field $|E(B=0)|$ is plotted as a
function of the magnetic field, expressed in units of the exchange
field $B_{\mathrm{ex}}$.

The magnetic field dependence of the polaron relaxation time
calculated after Eqs.~\eqref{t12} and ~\eqref{t10} is shown in
Fig.~\ref{fig:6}(b). In order to have it in a universal form and
highlight the two regimes, the results are presented by using
normalized values, namely $\ln(\tau_{\mathrm{MP}}/\tau_0)$ is
plotted versus $B/B_{\mathrm{ex}}$. Here, the parameter of the shown
dependences is the ratio of the HMP energy at zero field to
$k_\mathrm{B}T$. A larger $E(B=0)$ corresponds to a higher potential
barrier between the two stable HMP states, and correspondingly to a
longer spin relaxation time.

It is insightful to relate the polaron parameters determined by the
fit of the magnetic field dependence of the MP spin relaxation time
to the MP binding energy usually measured in photoluminescence
spectra under selective excitation. As found in
Ref.~\onlinecite{Merk95}, the MP is destroyed by an in-plane
magnetic field when its strength becomes larger than the critical
field $B_3$ given by Eq.(15) of this paper. In our notation,
\begin{equation}
B_3=B_{\mathrm{ex}}\frac{g_{zz}}{g_{\bot}}. \label{B3}
\end{equation}
One can see that the magnetic fields applied in our experiment are
well below $B_3$, and, therefore, the MP binding energy should
approximately be equal to its value at zero magnetic field
\begin{equation}
E_{\mathrm{MP}}(B)\approx E_{\mathrm{MP}}(B=0)=\frac{1}{2}\frac{d \Delta E^{\mathrm{hh}}_{z}}{dB}B_{\mathrm{ex}}=|E(B=0)|.
\label{EMP}
\end{equation}

\section{Discussion}
\label{sec:5}

Let us apply the developed model to the experimental data on the HMP
spin relaxation. A strong dependence of $\tau_{\mathrm{MP}}$ on the
magnetic field has been found experimentally, see Fig.~\ref{fig:4}.
The line shows the best fit to the data using Eqs.~\eqref{t12} and
~\eqref{t10}, for which the value $d \Delta E^{\mathrm{hh}}_{z}/dB =
21.6$~meV/T was taken from the measured electron spin beats in the
inset of Fig.~\ref{fig:B}. The two fitting parameters
$B_{\mathrm{ex}}$ and $g_{\bot}/g_{zz}$ are in fact independent, as
they are linked to the characteristic field, where the slope
changes, and to the slope in high magnetic fields, respectively.
$g_{\bot}/g_{zz}=0.04$ is in good agreement with the expectation of
a strongly anisotropic heavy-hole spin state in narrow QWs.
$B_{\mathrm{ex}}=0.05$~T corresponds to a binding energy of the HMP
of $E_{\mathrm{MP}}=|E(B=0)|=0.5$~meV for the QW with $x=0.04$.
These values are reasonable for QWs with low Mn
concentrations~\cite{Merk, Kusrayev08}.

The temperature dependence of $\tau_{\mathrm{MP}}$ shown in
Fig.~\ref{fig:2}(c) is in agreement with the HMP energy evaluated
from its magnetic field dependence. Here, the parameters of the
linear interpolations shown by the lines are $y=3.0 exp(x/0.304)$
for $x=0.04$ and $ y=2.5 exp(x/0.606)$ for $x=0.02$. This means
that at $B=0.5$~T, $\tau_0=3$ and 2.5~ns in the samples with
$x=0.04$ and 0.02, respectively. This time $\tau_0$ is a pre-factor,
which is of the order of the hole spin-flip time.

The approach developed in Ref.~\onlinecite{Kavokin99b} allows us to
evaluate the volume of the HMP. For that, we use Eq.~\eqref{t20} with
the known parameters for (Cd,Mn)Te: $\beta N_0=880$~meV and
$N_0^{-1}=0.056$~nm$^3$ \cite{Furdyna88}. For the 4-nm-thick QW with
$x=0.04$, where $B_{\mathrm{ex}}=0.05$~T, the estimated HMP volume
is $V_{\mathrm{MP}}=4300$~nm$^3$. Accordingly, the localization
radius of the resident hole is about 19~nm, which is reasonable for
the QW width fluctuations.

The mechanism of optical orientation of equilibrium magnetic
polarons in DMS QWs suggested in this paper has not been reported so
far. Here, it has been demonstrated experimentally for hole MPs, but
it is equally valid for electron MPs, as well as for donor-bound and
acceptor-bound MPs. The mechanism has some similarity with the
mechanism for spin coherence generation in electron- and hole-gases
of low density in nonmagnetic QWs and in singly-charged quantum dots
under resonant trion excitation. It has been validated for GaAs-,
(In,Ga)As-, CdTe- and ZnSe-based QWs and (In,Ga)As-based quantum
dots, for which charged exciton (trion) states are of key importance
for the optical properties \cite{Yakovlev08,Zh07,Fokina10,Zh14}.

The suggested mechanism for equilibrium magnetic polarons is
fundamentally different from the one for optical orientation of
exciton magnetic polarons (EMP) that was reported for bulk (Cd,Mn)Se
and for (Cd,Mn)Te- and (Cd,Mn)Te-based QWs
\cite{Warnock1,Warnock2,Merkulov95}. There are several aspects to
distinguish:

(i) For exciton MPs optical excitation was performed in the tail of
localized exciton states and the degree of induced optical
orientation is controlled by the magnetic fluctuations of Mn spins,
$\mathbf{M}_f$, in the exciton volume, prior the evolution of the
collective polaron. In the suggested mechanism, $\mathbf{M}_f$ does
not play a key role for the generation efficiency.

(ii) The EMP lifetime is limited by exciton recombination (typically
in the range of $30-500$~ps in direct band gap semiconductors). As a
result, EMP cannot always reach their equilibrium polaron energy.
Also, EMPs cannot be considered as suitable candidates for a
long-term spin memory. Contrary to that, the equilibrium MPs formed
from resident holes (or electrons) do not suffer from recombination;
as they are in equilibrium, their energy is not influenced by
dynamical factors.

(iii) In the case of the orientation of the equilibrium MPs via resonant
excitation of the underlying trion state we do not induce a
perturbation of the equilibrium MPs whose dynamics are in focus of our measurements. Instead, we rearrange the orientation of the other polarons
in the ensemble. Due to that we have the possibility to study the
properties of the equilibrium MP state, which remains unperturbed by
the optical excitation. This is as well different from the EMP
optical orientation, where the oriented MPs are formed from the
photogenerated excitons.

\section{Conclusions}
\label{sec:6}

We report on the observation of equilibrium hole magnetic polarons
in (Cd,Mn)Te/(Cd,Mn,Mg)Te QWs. Their optical orientation is provided
by resonant excitation of the trion states in these structures. The
long spin relaxation times of the magnetic polarons up to 60~ns are
provided by the anisotropic spin state of the heavy holes in the
QWs. These times may be further extended in structures with higher
Mn concentrations and enhanced carrier localization, which result in
larger polaron energies. The suggested mechanism of optical
orientation allows for the study of the polaron spin dynamics by
time-resolved pump-probe Kerr rotation. The mechanism is valid for
both the collective and fluctuation regimes of hole or electron MPs
and also for bound magnetic polarons, where the resident carrier is
bound either to a donor (electron) or an acceptor (hole).

{\bf Acknowledgements} We acknowledge the financial support by the
Russian Science Foundation (Grant No. 14-42-00015) and the Deutsche
Forschungsgemeinschaft in the frame of ICRC TRR 160. The research in Poland was partially supported by the National Science Centre (Poland) through Grants No. DEC-2012/06/A/ST3/00247 and No. DEC-2014/14/M/ST3/00484 and by   the Foundation for Polish Science through the Master program (G.K.).


\begin{thebibliography}{}
\bibitem{Comm1} We do not consider here effects related to the mixing of
heavy- and light-hole states with increasing magnetic field and the appearance
of nonzero in-plane components of the heavy-hole spin. While their
presence would not change the basics of the suggested approach and
can be accounted by its elaboration.
\end{thebibliography}
\end{document}